# Mass Pattern of the SM Fermions: Flavor Democracy Revisited

U. Kaya[§]
Ankara University, Ankara, Turkey
umit.kaya@cern.ch

S. Sultansoy
TOBB University of Economics and Technology, Ankara, Turkey
ANAS Institute of Physics, Baku, Azerbaijan
ssultansoy@etu.edu.tr

**Abstract**

Mass pattern of the SM fermions is one of the most important mysterious in particle physics. Flavor Democracy could shed light on this mystery. Addition of isosinglet quark and isosinglet lepton give opportunity to obtain masses of charged leptons and quarks of the 2$^{nd}$ and 3$^{rd}$ family due to small deviations of full Flavor Democracy.

[§] Corresponding Author

## 1. Introduction

Mass and mixing patterns of the SM fermions are among the most important issues, which should be clarified in particle physics. In recent interview published in CERN Courier [1], Steven Weinberg emphasized this point: "Asked what single mystery, if he could choose, he would like to see solved in his lifetime, Weinberg doesn't have to think for long: he wants to be able to explain the observed pattern of quark and lepton masses". In our opinion, flavor democracy (see reviews [2-5] and references therein) could provide an important key to solve this mystery.

In this paper we deal with mass pattern of the SM fermions (mixing pattern will be considered separately). In Section 2, mass pattern of the SM fermions are summarized. Section 3 is devoted to the current status of the Chiral Fourth Family. Possible solution of the mass pattern mystery due to adding new isosinglet down quark and charged lepton has been considered in Section 4. Finally, we summarized our results in Section 5.

## 2. Mass Pattern of the SM Fermions

Masses of known charged leptons and quarks are given in Table 1 [6]. We do not consider neutrino masses, since their values are not assigned experimentally and probably have specific mechanism, namely, see-saw. It should be noted that right-handed components of neutrinos are counterparts of the right-handed components of up quarks, therefore their inclusion does not mean BSM.

It is seen from Table 1 that masses of the first family fermions are much less than masses of second family ones and the latters are much lighter than the masses of the third family fermions (fermion mass hierarchy). Second important point is that mass of t quark is much greater than masses of tau lepton and b quark. This point exclude Flavor Democracy for 3 SM family case.

Table 1. Mass pattern of charged leptons and quarks.

|  | charged leptons | Up type quarks | Down type quarks |
|---|---|---|---|
| **1st Family** | $0.510998928 \pm 1.1 \times 10^{-8}$ MeV | $2.3 ^{+0.7}_{-0.5}$ MeV | $4.8 ^{+0.5}_{-0.3}$ GeV |
| **2nd Family** | $105.6583715 \pm 3.5 \times 10^{-6}$ MeV | $1.275 \pm 0.025$ GeV | $95 \pm 5$ MeV |
| **3rd Family** | $1776.82 \pm 0.16$ MeV | $173.21 \pm 0.51 \pm 0.71$ GeV | $4.18 \pm 0.03$ GeV |

## 3. Status of the Chiral Fourth Family (C4F)

It is known that the Standard Model does not fix the number of fermion families. This number should be less than 9 in order to preserve asymptotic freedom and more than 2 in order to provide CP violation. According to the LEP data on Z decays, number of chiral families with light neutrinos ($m_\nu \ll m_Z$) is equal to 3, whereas extra families with heavy neutrinos are not forbidden. The fourth chiral family was widely discussed thirty years ago (see, for example [7, 8]). However, the topic was pushed off the agenda due to the misinterpretation of the LEP data.

Twenty years later 3 workshops on the fourth SM family [9–11] were held (for summary of the first and third workshops see [12] and [13], respectively). Main motivation was Flavor Democracy [14–16], which naturally provides heavy fourth family fermions including neutrino (consequences of Flavor Democracy Hypothesis for different models, including MSSM and E6, have been considered in [17, 18]). In addition, fourth family gives opportunity to explain baryon asymmetry of universe; it can accommodate emerging possible hints of new physics in rare decays of heavy mesons etc (see [12] and references therein). Phenomenological papers on direct production (including anomalous resonant production) of the SM4 fermions at different colliders are reviewed in [19] (see tables VI and VII in [19]).

This activity has almost ended due to misinterpretation of the LHC data on the Higgs decays. It should be emphasized that these data exclude the minimal SM4 with one Higgs doublet, whereas non-minimal SM4 with extended Higgs sector are still allowed [20, 21]. On the other hand, partial wave unitarity puts an upper limit around 700 GeV on the masses of fourth SM family quarks [22], which is almost excluded by the recent ATLAS and CMS data on search for pair production. For example, ATLAS $\sqrt{s} = 8$ TeV data with 20.3 fb$^{-1}$ integrated luminosity excludes new chiral quarks with mass below 690 GeV at 95% confidence level assuming BR(Q → W q)=1 [23].

Even if SM4 may be excluded by the LHC soon, this is not the case for the general chiral fourth family (C4F). Therefore, ATLAS and CMS should continue a search for C4F up to kinematical limits. Concerning pair production, rescaling of the ATLAS lower bound using collider reach framework [24] shows that LHC with opportunity to cover Mu4 up to 0.94, 1.25, 1.50, and 2.13 TeV with integrated luminosities 20, 100, 300 and 3000 fb$^{-1}$, respectively.

## 4. New Weak Isosinglet Fermions

As mentioned in [2, 3, 5], large mass difference between $m_b$ and $m_t$ can be explained by the addition of isosinglet quarks. Here we consider an addition of one isosinglet quark, so the quark sector is determined as

$$\begin{pmatrix} u_L \\ d_L \end{pmatrix}, \begin{pmatrix} c_L \\ s_L \end{pmatrix}, \begin{pmatrix} t_L \\ b_L \end{pmatrix}, u_R\ d_R,\ c_R\ s_R,\ t_R\ b_R,\ D_L\ D_R \quad (1)$$

where D denotes new isosinglet quark.

In the case of full flavor democracy the mass matrix of the up type quarks can be written as

|     | $u_R$ | $c_L$ | $t_L$ |
|-----|-------|-------|-------|
| $u_L$ | $a\eta$ | $a\eta$ | $a\eta$ |
| $c_L$ | $a\eta$ | $a\eta$ | $a\eta$ |
| $t_L$ | $a\eta$ | $a\eta$ | $a\eta$ |

(2)

and mass matrix of down type quarks is

|       | $d_R$ | $s_R$ | $b_R$ | $D_R$ |
|-------|-------|-------|-------|-------|
| $d_L$ | $a\eta$ | $a\eta$ | $a\eta$ | $a\eta$ |
| $s_L$ | $a\eta$ | $a\eta$ | $a\eta$ | $a\eta$ |
| $b_L$ | $a\eta$ | $a\eta$ | $a\eta$ | $a\eta$ |
| $D_L$ | $M$ | $M$ | $M$ | $M$ |

(3)

where $M$ ($M \gg \eta$) is the new physics scale that determines the mass of isosinglet quark. In this case $m_u = m_c = 0$ and $m_t = 3a\eta$ for up type quarks, $m_d = m_s = m_b = 0$ and $m_D = 3a\eta + M = m_t + M$ for down type quarks.

In order to obtain mass of b quark, small deviation from matrix (3) is involved, namely

|       | $d_R$ | $s_R$ | $b_R$ | $D_R$ |
|-------|-------|-------|-------|-------|
| $d_L$ | $a\eta$ | $a\eta$ | $a\eta$ | $(1-\alpha_b)a\eta$ |
| $s_L$ | $a\eta$ | $a\eta$ | $a\eta$ | $(1-\alpha_b)a\eta$ |
| $b_L$ | $a\eta$ | $a\eta$ | $a\eta$ | $(1-\alpha_b)a\eta$ |
| $D_L$ | $(1-\beta_b)M$ | $(1-\beta_b)M$ | $(1-\beta_b)M$ | $M$ |

(4)

At this stage, for numerical calculations we assume $\alpha_b = \beta_b$. The masses of d and s quarks remain as $m_d = m_s = 0$. On the other hand masses of b and D quarks are as follows

$$m_b = \frac{1}{2}\{3a\eta + M - \sqrt{...}\}; \quad m_D = \frac{1}{2}\{3a\eta + M + \sqrt{...}\}$$
$$\sqrt{...} = \sqrt{9(a\eta)^2 + 12a\eta M \alpha_b \beta_b - 12a\eta M \alpha_b - 12a\eta M \beta_b + 6a\eta M + M^2} \quad (5)$$

For $\alpha_b = \beta_b \ll 1$

$$\sqrt{...} = (3a\eta + M) - \frac{12a\eta M \alpha_b}{(3a\eta+M)} + \frac{6a\eta M \alpha_b^2 (M-3a\eta)^2 (3a\eta+M)}{(3a\eta+M)^4} + \cdots \quad (6)$$

Therefore $m_b$, $m_D$ and $\alpha_b$ are given by

$$m_b = \frac{6a\eta M \alpha_b}{(3a\eta+M)} - \frac{3a\eta M \alpha_b^2 (M-3a\eta)^2 (3a\eta+M)}{(3a\eta+M)^4} \quad (7)$$

$$m_D = (3a\eta + M) - \frac{6a\eta M \alpha_b}{(3a\eta+M)} + \frac{3a\eta M \alpha_b^2 (M-3a\eta)^2 (3a\eta+M)}{(3a\eta+M)^4} \quad (8)$$

$$\alpha_b = \frac{1}{6a\eta M (M-3a\eta)^2}\{27(a\eta)^3 + \sqrt{3}\sqrt{a\eta M(3a\eta+M)^3 9(a\eta)^2(M-2m_b) + 3a\eta(4m_b+M) - 2m_b M^2} + 18(a\eta)^2 M^2 + 3a\eta M^2\} \quad (9)$$

Taking $m_b = 4.18\ GeV$ and $m_t = 173\ GeV$ we obtain $\alpha_b$ and $m_D$ corresponding to the different values of M which are given in Table 2.

Table 2. $\alpha_b$ and $m_D$ correspoding to different values of M.

| M (GeV) | 1000 | 2000 | 5000 | 10000 | 20000 |
|---|---|---|---|---|---|
| $\alpha_b$ $(10^{-2})$ | 1.43 | 1.33 | 1.26 | 1.24 | 1.23 |
| $m_D$ | 1168 | 2168 | 5168 | 10168 | 20168 |
| $\alpha_\tau$ $(10^{-3})$ | 6.04 | 5.60 | 5.34 | 5.25 | 5.21 |
| $m_L$ | 1171 | 2171 | 5171 | 10171 | 20171 |

Similarly, tau lepton mass can be determined by adding an isosinglet lepton. In this case lepton sector is

$$\begin{pmatrix} e_L \\ \nu_{e_L} \end{pmatrix}, \begin{pmatrix} \mu_L \\ \nu_{\mu_L} \end{pmatrix}, \begin{pmatrix} \tau_L \\ \nu_{\tau_L} \end{pmatrix}, e_R\ \nu_{e_R}, \mu_R\ \nu_{\mu_R}, \tau_R\ \nu_{\tau_R}, L_L\ L_R \qquad (10)$$

Then, the mass matrix becomes

$$\begin{array}{c|cccc} & e_R & \mu_R & \tau_R & L_R \\ e_L & a\eta & a\eta & a\eta & (1-\alpha_\tau)a\eta \\ \mu_L & a\eta & a\eta & a\eta & (1-\alpha_\tau)a\eta \\ \tau_L & a\eta & a\eta & a\eta & (1-\alpha_\tau)a\eta \\ L_L & (1-\beta_\tau)M & (1-\beta_\tau)M & (1-\beta_\tau)M & M \end{array} \qquad (11)$$

For $\alpha_\tau = \beta_\tau \ll 1$  $m_\tau, m_L$ and $\alpha_\tau$ are given by

$$m_\tau = \frac{6a\eta M \alpha_\tau}{(3a\eta+M)} - \frac{3a\eta M \alpha_\tau^2 (M-3a\eta)^2(3a\eta+M)}{(3a\eta+M)^4} \qquad (12)$$

$$m_L = (3a\eta + M) - \frac{6a\eta M \alpha_\tau}{(3a\eta+M)} + \frac{3a\eta M \alpha_\tau^2 (M-3a\eta)^2(3a\eta+M)}{(3a\eta+M)^4} \qquad (13)$$

$$\alpha_\tau = \frac{1}{6a\eta M(M-3a\eta)^2}\{27(a\eta)^3 +$$
$$\sqrt{3}\sqrt{a\eta M(3a\eta+M)^3} 9(a\eta)^2(M-2m_\tau) + 3a\eta(4m_\tau + M) - 2m_\tau M^2 +$$
$$18(a\eta)^2 M^2 + 3a\eta M^2\} \qquad (14)$$

With $m_\tau = 1.777\ GeV$ we obtain $\alpha_\tau$ and $m_L$ corresponding to the different values of M which are given in the last two rows of Table 2.

Because the masses of the e, u and d quarks are very small, we do not comment on them at this stage. Masses of s quark, muon and c quark can also be obtained due to small deviations from full democracy. Concerning c quark let us consider following modification of the mass matrix of up quarks

$$\begin{array}{c|ccc} & u_R & c_L & t_L \\ u_L & a\eta & a\eta & a\eta \\ c_L & a\eta & a\eta & a\eta \\ t_L & a\eta & a\eta & (1+\alpha_c)a\eta \end{array} \qquad (15)$$

For $\alpha_c \ll 1$

$$m_u = 0; \; m_c = \frac{1}{2}a\eta\{\alpha_c + 3 - \sqrt{...}\}; \; m_t = \frac{1}{2}a\eta\{\alpha_c + 3 - \sqrt{...}\} \tag{16}$$

Therefore

$$\alpha_c \cong \frac{9m_c}{2m_t} = 3.3\times10^{-2} \tag{17}$$

In order to obtain muon mass, we consider following modification of the Eq. 4

$$
\begin{array}{c|cccc}
 & e_R & \mu_R & \tau_R & L_R \\
e_L & a\eta & a\eta & a\eta & (1-\alpha_\tau)a\eta \\
\mu_L & a\eta & a\eta & a\eta & (1-\alpha_\tau)a\eta \\
\tau_L & a\eta & a\eta & (1+\alpha_\mu)a\eta & (1-\alpha_\tau)a\eta \\
L_L & (1-\beta_\tau)M & (1-\beta_\tau)M & (1-\beta_\tau)M & M
\end{array}
\tag{18}
$$

For $M = 2000\ GeV, \alpha_\tau = \beta_\tau = 5.58\times10^{-3}$ and $\alpha_\mu = 2.73\times10^{-4}$ this mass matrix lead to

$$m_L = 2171\ GeV, m_\tau = 1.777\ GeV, m_\mu = 104.7\ MeV \tag{19}$$

Similarly for down type quarks

$$
\begin{array}{c|cccc}
 & d_R & s_R & b_R & D_R \\
d_L & a\eta & a\eta & a\eta & (1-\alpha_b)a\eta \\
s_L & a\eta & a\eta & a\eta & (1-\alpha_b)a\eta \\
b_L & a\eta & a\eta & (1+\alpha_c)a\eta & (1-\alpha_b)a\eta \\
D_L & (1-\beta_b)M & (1-\beta_b)M & (1-\beta_b)M & M
\end{array}
\tag{20}
$$

For $M = 2000\ GeV, \alpha_b = \beta_b = 1.32\times10^{-2}$ and $\alpha_c = 2.48\times10^{-4}$ we obtain

$$m_D = 2168\ GeV, m_b = 4.18\ GeV, m_s = 95.2\ MeV \tag{21}$$

## 5. Conclusion

It is shown that masses of 2nd and 3rd SM family fermions can be obtained due to small deviations of flavor democracy, if new heavy isosinglet quark and isosinglet lepton exist in nature. These new quark and lepton have approximately same masses. Main decay channels of isosinglet quarks are $D \rightarrow Wt$ with $BR\sim0.5$, $D \rightarrow Zb$ with $BR\sim0.25$ and $D \rightarrow Hb$ with $BR\sim0.25$. Isosinglet lepton will decay into $L \rightarrow W\nu$ with $BR\sim0.5$, $L \rightarrow Z\tau$ with $BR\sim0.25$ and $L \rightarrow H\tau$ with $BR\sim0.25$. According to the ATLAS experiment results $m_D > 1.25\ TeV$ for $D \rightarrow Wt$ channel [25].

**Reference:**

[1] http://cerncourier.com/cws/article/cern/70138

[2] S. Sultansoy, Four Ways to TeV Scale, Turk.J.Phys. 22 (1998) 575-594.

http://inspirehep.net/record/440270

[3] S. Sultansoy, Four Remarks on Physics at LHC, Invited talk at ATLAS Week, 26-31 May 1997. Geneva, Switzerland.

http://inspirehep.net/record/971257

[4] E. Arik and S. Sultansoy, Turkish comments on 'Future perspectives in HEP'.

http://inspirehep.net/record/612725

[5] S. Sultansoy, Flavor Democracy in Particle Physics.

http://inspirehep.net/record/729701

[6] C. Patrignani et al. (Particle Data Group), Chin. Phys. C, 40, 100001 (2016) and 2017 update.

http://pdg.lbl.gov

[7] Proceedings of the First International Symposium on the fourth family of quarks and leptons, 26-28 February 1987, Santa Monica, CA, published in Annals of the New York Academy of Sciences, 518 (1987), edited by D. Cline and A. Soni.

[8] Proceedings of the Second International Symposium on the fourth family of quarks and leptons, 23-25 February 1989, Santa Monica, CA, published in Annals of the New York Academy of Sciences, 578 (1989), edited by D. Cline and A. Soni.

[9] Beyond the 3SM generation at the LHC era Workshop, 4-5 September 2008, CERN, Geneva, Switzerland.

http://indico.cern.ch/event/33285

[10] Second Workshop on Beyond 3 Generation Standard Model New Fermions at the Crossroads of Tevatron and LHC, 14-16 January 2010 Taipei, Taiwan.

http://indico.cern.ch/event/68036

[11] Third Workshop on Beyond 3 Generation Standard Model Under the light of the initial LHC results, 23-25 October 2011, Istanbul, Turkey.

https://indico.cern.ch/event/150154

[12] B. Holdom et al., "Four statements about the fourth generation", PMC Physics A 2009, 3:4; arXiv:0904.4698 [hep-ph].

[13] S. A. Cetin et al., "Status of the Fourth Generation: A Brief Summary of B3SM-III Workshop in Four Parts", arXiv:1112.2907.

[14] H. Fritzsch, "Light neutrinos, nonuniversality of the leptonic weak interaction and a fourth massive generation", Phys. Lett. B 289.1-2 (1992) 92-96.

[15] A. Datta, "Flavour democracy calls for the fourth generation", Pramana 40.6 (1993): L503-L509.

[16] A. Celikel, A. K. Ciftci, and S. Sultansoy, "A search for the fourth SM family", Phys. Lett. B 342.1 (1995) 257-261.

[17] S. Sultansoy, "Why the four SM families", arXiv:hep-ph/0004271.

[18] S. Sultansoy, "Flavor democracy in particle physics", AIP Conf. Proc. 899 (2007) 49; arXiv:hep-ph/0610279.

[19] M. Sahin, S. Sultansoy, and S. Turkoz, "Search for the fourth standard model family", Phys. Rev. D 83 (2011) 054022; arXiv:1009.5405v2.

[20] S. Bar-Shalom, M. Geller, S. Nandi and A. Soni, "Two Higgs doublets, a 4th generation and a 125 GeV Higgs: a review", Adv. High Energy Phys. (2013) 672972, arXiv:1208.3195 [hep-ph]

[21] S. Banerjee, M. Frank, S. K. Rai, "Higgs data confronts sequential fourth generation fermions in the Higgs triplet model", Phys.Rev. D89 (2014) no.7, 075005.

[22] M. S. Chanowitz, M. A. Furman and I. Hinchliffe, "Weak interactions of ultra heavy fermions (II)", Nucl. Phys. B153 (1979) 402.

[23] G. Aad et al., "Search for pair production of a new heavy quark that decays into a W boson and a light quark in p p collisions at s= 8 TeV with the ATLAS detector", Phys. Rev. D 92.11 (2015) 112007.

[24] G. Salam and A. Weiler, "The Collider Reach project".
http://collider-reach.web.cern.ch/collider-reach

[25] ATLAS Exotics Searching Summary,
https://atlas.web.cern.ch/Atlas/GROUPS/PHYSICS/CombinedSummaryPlots/EXOTICS/ATLAS_Exotics_Summary/ATLAS_Exotics_Summary.png